\documentclass[useAMS,usenatbib,usegraphicx]{mn2e}
\usepackage{hyperref}
\newcommand*{\figref}[1]{Figure~\ref{#1}}

\newcommand*{\tabref}[1]{Table~\ref{#1}}
\newcommand*{\eqnref}[1]{Equation~\ref{#1}}
\newcommand*{\secref}[1]{Section~\ref{#1}}

\newcommand*{\unit}[1]{\ensuremath{\mathrm{\, #1}}}
\newcommand*{\Msun}{\unit{M_{\odot}}}
\newcommand*{\erg}{\unit{erg}}
\newcommand*{\keV}{\unit{keV}}
\newcommand*{\second}{\unit{s}}
\newcommand*{\cm}{\unit{cm}}
\newcommand*{\km}{\unit{km}}
\newcommand*{\Mpc}{\unit{Mpc}}

\newcommand*{\E}[1]{\ensuremath{\times 10^{#1}}}
\newcommand*{\chisq}{\ensuremath{\chi^2}}

\newcommand*{\mysub}[2]{\ensuremath{#1_{\mathrm{#2}}}}
\newcommand*{\Omegam}{\mysub{\Omega}{m}}
\newcommand*{\Omegab}{\mysub{\Omega}{b}}
\newcommand*{\Omegac}{\mysub{\Omega}{c}}

\newcommand*{\LCDM}{\ensuremath{\Lambda}CDM}
\newcommand*{\wCDM}{\ensuremath{w}CDM}

\newcommand*{\JMF}{J01}
\newcommand*{\RB}{RB02}

\newcommand*{\Chandra}{{\it{Chandra}}}
\newcommand*{\ROSAT}{{\it{ROSAT}}}
\newcommand*{\fsky}{\mysub{f}{sky}}
\newcommand*{\modchisq}{\ensuremath{\tilde{\chi}^2}}
\newcommand*{\llxlf}{\mysub{\ell}{XLF}}
\newcommand*{\ie}{\emph{i.e.}}
\newcommand*{\cf}{\emph{cf.}}
\newcommand*{\fgas}{\mysub{f}{gas}}

\newcommand*{\iscatsym}{\ensuremath{\eta}}
\newcommand*{\iscat}{\ensuremath{\iscatsym_0}}
\newcommand*{\iscatprime}{\ensuremath{\iscatsym_1}}
\newcommand*{\mscat}{\ensuremath{\varepsilon}}
\newcommand*{\meanbias}{\ensuremath{\bar{b}}}
\newcommand*{\biasscatter}{\ensuremath{s_b}}

\title[New constraints on dark energy]{New constraints on dark energy from the observed growth of the most X-ray luminous galaxy clusters}
\author[A. Mantz et al.]{
  A.~Mantz,$^1$\thanks{E-mail: amantz@stanford.edu} S.~W.~Allen,$^1$ H.~Ebeling$^2$ and D.~Rapetti$^1$\\
  $^1$Kavli Institute for Particle Astrophysics and Cosmology, Stanford University, 382 Via Pueblo Mall, Stanford, CA 94305-4060, USA\\
  $^2$Institute for Astronomy, 2680 Woodlawn Drive, Honolulu, HI 96822, USA
}
\date{27 March 2008}

\begin{document}
\pagerange{\pageref{firstpage}--\pageref{lastpage}} \pubyear{2008}
\maketitle
\label{firstpage}

\begin{abstract}
  We present constraints on the mean matter density, \Omegam{}, the normalization of the density fluctuation power spectrum, $\sigma_8$, and the dark-energy equation-of-state parameter, $w$, obtained from measurements of the X-ray luminosity function of the largest known galaxy clusters at redshifts $z<0.7$, as compiled in the Massive Cluster Survey (MACS) and the local BCS and REFLEX galaxy cluster samples. Our analysis employs an observed mass--luminosity relation, calibrated by hydrodynamical simulations, including corrections for non-thermal pressure support and accounting for the presence of intrinsic scatter. Conservative allowances for all known systematic uncertainties are included, as are standard priors on the Hubble constant and mean baryon density. We find $\Omegam=0.28^{+0.11}_{-0.07}$ and $\sigma_8=0.78^{+0.11}_{-0.13}$ for a spatially flat, cosmological-constant model, and $\Omegam=0.24^{+0.15}_{-0.07}$, $\sigma_8=0.85^{+0.13}_{-0.20}$ and $w=-1.4^{+0.4}_{-0.7}$ for a flat, constant-$w$ model (marginalized 68 per cent confidence intervals). Our findings constitute the first determination of the dark-energy equation of state from measurements of the growth of cosmic structure in galaxy clusters, and the consistency of our result with $w=-1$ lends additional support to the cosmological-constant model. Future work improving our understanding of redshift evolution and observational biases affecting the mass--X-ray luminosity relation have the potential to significantly tighten these constraints. Our results are consistent with those from recent analyses of type Ia supernovae, cosmic microwave background anisotropies, the X-ray gas mass fraction of relaxed galaxy clusters, baryon acoustic oscillations and cosmic shear. Combining the new X-ray luminosity function data with current supernova, cosmic microwave background and cluster gas fraction data yields the improved constraints $\Omegam=0.269\pm 0.016$, $\sigma_8=0.82\pm 0.03$ and $w=-1.02\pm 0.06$.
\end{abstract}


\section{Introduction}
\label{sec:introduction}

In the hierarchical collapse scenario for structure formation in the universe, the number density of collapsed objects as a function of mass and cosmic time is a sensitive probe of cosmology. The galaxy clusters that occupy the high-mass tail of this population provide a powerful and relatively clean tool for cosmology, since their growth is predominantly determined by linear gravitational processes. In the past, the local population of galaxy clusters has been used to jointly constrain the average matter density of the universe and the amplitude of perturbations in the density field \citep[e.g.][]{Reiprich02,Seljak02,Viana02,Allen03,Pierpaoli03,Schuecker03,Voevodkin04,Dahle06,Rozo07}. Pushing observations to higher redshift breaks the degeneracy between those two parameters \citep[e.g.][]{Donahue99,Eke98,Henry00,Borgani01,Vikhlinin03}, and allows properties of dark energy to be probed as well \citep[e.g.][]{Haiman01,Levine02,Weller02,Majumdar03,Majumdar04,Henry04}.

Investigations of this type require sky surveys with well understood selection functions to find clusters, as well as a relation linking cluster mass with an observable. A successful solution to the former requirement has been to identify clusters by the X-ray emission produced by hot intracluster gas, notably using data from the \ROSAT{} All-Sky Survey \citep[RASS;][]{Trumper93}. The \ROSAT{} Brightest Cluster Sample \citep[BCS;][]{Ebeling98,Ebeling00} and \ROSAT-ESO Flux Limited X-ray sample \citep[REFLEX;][]{Bohringer04} together cover approximately two-thirds of the sky out to redshift $z\sim 0.3$ and contain more than 750 clusters. The Massive Cluster Survey \citep[MACS;][]{Ebeling01,Ebeling07} -- which at this writing contains 126 clusters and covers 55 per cent of the sky -- extends these data to $z\sim 0.7$.

The most straightforward mass--observable relation to complement these X-ray flux-limited surveys is the mass--X-ray luminosity relation. For sufficiently massive (hot) objects at the relevant redshifts, the conversion from X-ray flux to luminosity is approximately independent of temperature, in which case the luminosities can be estimated directly from the survey flux and the selection function is identical to the requirement of detection. In a flux-complete survey further restricted to high luminosities, every cluster should thus be usable in the analysis, without the need for additional observations other than those required to calibrate the mass--luminosity relation. A disadvantage is that there is a large scatter in cluster luminosities at fixed mass; however, sufficient data allow this scatter to be quantified empirically. Alternative approaches use cluster temperature \citep{Henry00,Seljak02,Pierpaoli03,Henry04}, gas fraction \citep{Voevodkin04} or $Y_X$ parameter \citep{Kravtsov06} to achieve tighter mass--observable relations at the expense of reducing the size of the samples available for analysis. The need to quantify the selection function in terms of both X-ray flux and a second observable additionally complicates these efforts.

In this paper, we use the observed X-ray luminosity function to investigate two cosmological scenarios, assuming a spatially flat metric in both cases: the first includes dark energy in the form of a cosmological constant (\LCDM{}); the second has dark energy with a constant equation-of-state parameter, $w$ (\wCDM{}). In the latter case, we account for the evolution of density perturbations in the dark-energy fluid, assuming that the dark-energy sound speed is equal to the speed of light. For each model, our results are in good agreement with findings from independent cosmological data sets, notably type Ia supernovae (SNIa), the cosmic microwave background (CMB), the X-ray gas mass fraction of galaxy clusters (\fgas{}), and measurements of cosmic shear.

The theoretical background for this work is reviewed in \secref{sec:theory}. \secref{sec:observations} details the data used to constrain the mass--luminosity relation and the cluster samples used to measure the X-ray luminosity function. The analysis procedure is described in \secref{sec:analysis} and the cosmological results are presented in \secref{sec:results}. The importance of various systematic effects is discussed in \secref{sec:discussion}.

Unless otherwise noted, masses and luminosities quoted in this paper or shown in figures are computed with respect to a spatially flat \LCDM{} reference cosmology with Hubble constant $h=H_0/100\km\second^{-1}\Mpc^{-1}=0.7$ and $\Omegam=0.3$. Luminosities and fluxes refer specifically to the 0.1--2.4\keV{} energy band in the source and observer rest frames, respectively. We will consistently use the notation $L$ to denote to the true luminosity of a cluster and $\hat{L}$ to denote the luminosity inferred from observation. We will also write, for example, \Omegam{} to refer to the present day matter density in units of the critical density, whereas $\Omegam(z)$ is the same quantity at redshift $z$.

\section{Theory}
\label{sec:theory}

The variance of the linearly evolved density field, smoothed by a spherical top-hat window of comoving radius $R$, enclosing mass $M=4\pi \bar{\rho}_m R^3 / 3$, is
\begin{equation}
  \sigma^2(M,z) = \frac{D^2(z)}{2\pi^2} \int_0^\infty k^2 P(k) |W_M(k)|^2 dk.
\end{equation}
Here $\bar{\rho}_m$ is the mean comoving matter density of the universe, $P(k)$ is the linear power spectrum at redshift zero, $W_M(k)$ is the Fourier transform of the window function, and $D(z)=\sigma_8(z)/\sigma_8(0)$ is the growth factor of linear perturbations at scales of $8h^{-1}\Mpc$, normalized to unity at $z=0$. We evaluate the transfer function, $T(k)$, and power spectrum, $P(k)\propto k^{n_s} T^2(k)$, using the CAMB package of \citet{Lewis00}.\footnote{http://www.camb.info/}

\citet[][hereafter \JMF{}]{Jenkins01} and \citet{Evrard02} have shown that the predicted mass function of galaxy clusters of mass $M$ at redshift $z$ can be written in terms of a ``universal'' function of $\sigma^{-1}(M,z)$, 
\begin{equation}
  f(\sigma^{-1}) = \frac{M}{\bar{\rho}_m} \frac{dn(M,z)}{d\ln\sigma^{-1}},
\end{equation}
which can be fit by a simple form,
\begin{equation}
  \label{eq:jmf}
  f(\sigma^{-1}) = A \exp \left( -|\ln \sigma^{-1}+B|^\epsilon  \right),
\end{equation}
for cosmological-constant models. It has since been verified that this fitting function is approximately universal (within $\sim 20$ per cent) among models with constant $w\neq-1$ and some evolving-$w$ models \citep{Klypin03,Linder03,Lokas04,Kuhlen05}. We adopt the values $A=0.316$, $B=0.67$ and $\epsilon=3.82$ from \JMF{}, determined using a spherical overdensity group finder at an overdensity of 324 with respect to the mean matter density ($324m$). The number density per unit mass of galaxy clusters of mass $M$ at redshift $z$ is then
\begin{equation}
  \label{eq:mfcn_def}
  \frac{dn(M,z)}{dM} = \frac{\bar{\rho}_m}{M} \frac{d\ln \sigma^{-1}}{dM} f(\sigma^{-1}).
\end{equation}

Following \citet{Morandi07}, we describe the relationship between mass and X-ray luminosity for massive clusters as self-similar \citep[e.g.][]{Bryan98}, modified by an additional redshift-dependent factor
\begin{equation}
  \label{eq:MLpowlaw}
  E(z)M_\Delta = M_0 \left(\frac{L}{E(z)}\right)^\beta (1+z)^\gamma,
\end{equation}
where $E(z)=H(z)/H_0$ and $M_\Delta$ is the cluster mass defined at an overdensity of $\Delta$ with respect to the critical density (\ie{} where the edge of the cluster is defined such that the mean density enclosed is a fixed factor $\Delta$ times the critical density). Our model also includes a log-normal scatter of width $\iscatsym(z)$ in luminosity for a given mass. There is good theoretical reason to expect that this scatter may evolve with redshift, since some of the variability in cluster luminosities can be attributed to the development of cool cores as well as the influence of  merger events \citep{OHara06,Chen07}. Since current data are insufficient to directly detect evolution in the mass--luminosity scatter (\citealt{OHara07} and references therein), we allow for a generic, linear evolution in the scatter:
\begin{equation}
  \label{eq:MLsevol}
  \iscatsym(z) = \iscat \left( 1 + \iscatprime z \right).
\end{equation}

The form of the self-similar evolution appearing in \eqnref{eq:MLpowlaw} is appropriate for constraining the mass--luminosity relation at fixed overdensity with respect to the critical density. We have chosen to constrain the relation at an overdensity of $\Delta=500$ in order to match our data (\secref{sec:RBdata}), and because the simulations of \citet{Evrard07} indicate that the virial relations have lower scatter at this overdensity compared with higher overdensities. The fact that the mass function and mass--luminosity relation use different definitions of cluster mass requires us to convert masses from one overdensity to another during the analysis \citep{White02}; details of carrying out this conversion, assuming a spherically symmetric \citet*{Navarro97} density profile, are reviewed in \citet{Hu03}.

Equations \ref{eq:mfcn_def}--\ref{eq:MLsevol}, combined with a suitable stochastic model linking the true luminosity of a cluster with its observed flux, provide the means to predict the galaxy cluster luminosity function based on a set of model parameters (\secref{sec:analysis}).

\section{Observations}
\label{sec:observations}

\subsection{Mass--luminosity relation}
\label{sec:MLrelation}

\subsubsection{X-ray data}
\label{sec:RBdata}

\begin{figure}
  \center
  \includegraphics[angle=270]{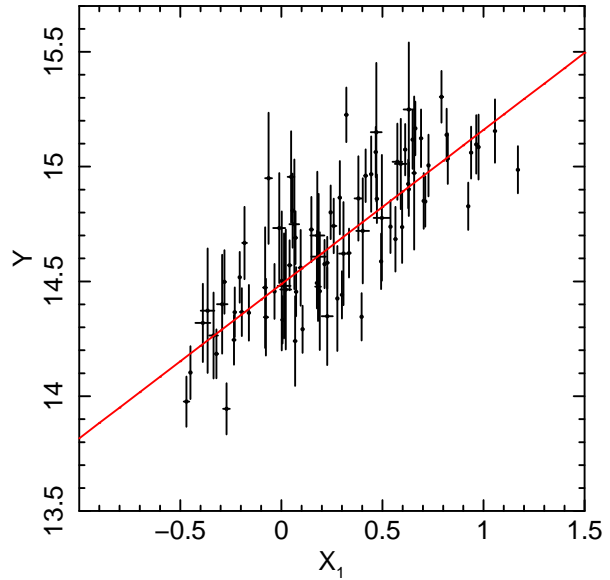}
  \caption{Mass--luminosity data of \RB{} and the best-fitting relation (\eqnref{eq:MLmodel}). The quantities $Y$ and $X_1$ are defined in \eqnref{eq:MLmodel2} and relate to log-mass and log-luminosity, respectively. As noted in the text, objects with $X_1<-0.5$ or $z>0.11$ are excluded from the analysis. All masses and luminosities are computed for our reference cosmology.}
  \label{fig:ML_x1_y}
\end{figure}

We have determined the mass--luminosity relation of galaxy clusters at low redshift using the sample of \citet[][hereafter \RB{}]{Reiprich02}, shown in \figref{fig:ML_x1_y}. In these data, the luminosities were measured from a combination of pointed \ROSAT{} PSPC and RASS data. The masses within $r_{500}$ ($M_{500}$) were determined by fitting the surface brightness profile with a $\beta$-model \citep{Cavaliere78} and applying the hydrostatic equation, assuming that the intracluster gas is isothermal. Here $r_{500}$ is the radius within which the mean density is 500 times the critical density, $\rho_c$ (not the universal mean matter density, $\bar{\rho}_m$).

The assumption of hydrostatic equilibrium introduces a bias into the derived masses due to the presence of non-thermal support and asphericity in the observed clusters. This bias has been observed in simulations \citep{Faltenbacher05,Rasia06,Nagai07} and in comparisons of masses derived from X-ray and weak gravitational lensing observations of the same targets \citep{Mahdavi07}, although the number of objects in these studies remains low. \citet{Rasia06} find a bias of $-34\pm14$ per cent at $r_{500}$ from simulations of 5 clusters. \citet{Nagai07} find a smaller bias, $-25\pm16$ per cent, from 16 simulated clusters at $z=0$. Note that there are differences in the methodology of the two groups; Rasia et al. use an isothermal $\beta$-model to determine the X-ray mass from mock images, while Nagai et al. use a more sophisticated X-ray analysis and additionally consider the observational bias in the determination of $r_{500}$. \citet{Mahdavi07} have compared masses determined from X-rays and lensing for 18 clusters and find a bias of $-22\pm9$ per cent at $r_{500}$, assuming the lensing masses are accurate. We adopt the Nagai et al. values, $-25\pm16$ per cent, as the nominal values of the mean bias, \meanbias{}, and scatter in bias between clusters, \biasscatter{}. In order to account for the spread among the various results for \meanbias{} and the small number of simulated clusters used to estimate \biasscatter{}, we assign a 20 per cent systematic uncertainty to both of these quantities (see below).

We fit the log-linear model (see \eqnref{eq:MLpowlaw})
\begin{equation}
  \label{eq:MLmodel}
  Y = \alpha + \beta X_1 + \gamma X_2,
\end{equation}
where
\begin{eqnarray}
  \label{eq:MLmodel2}
  Y & = & \log_{10}\left(\frac{E(z)M_{500}}{\Msun}\right), \nonumber \\
  X_1 & = & \log_{10}\left(\frac{L}{E(z)10^{44}\erg\second^{-1}}\right), \nonumber \\
  X_2 & = & \log_{10}\left(1+z\right), \nonumber \\
  \alpha & = & \log_{10}\left(\frac{M_0}{\Msun}\right).
\end{eqnarray}
The full \RB{} data set contains some small groups which can be influential in  the estimation of the slope, $\beta$. Since we only require the mass--luminosity relation to be calibrated approximately one decade below the luminosity cut used for the X-ray luminosity function data (\secref{sec:XLF}), we fit only the data with $X_1>-0.5$ in our reference cosmology. This choice mitigates the possibility of obtaining biased results if slope of the mass--luminosity relation is different for massive clusters compared with smaller groups.

\begin{figure}
  \center
  \includegraphics[angle=270]{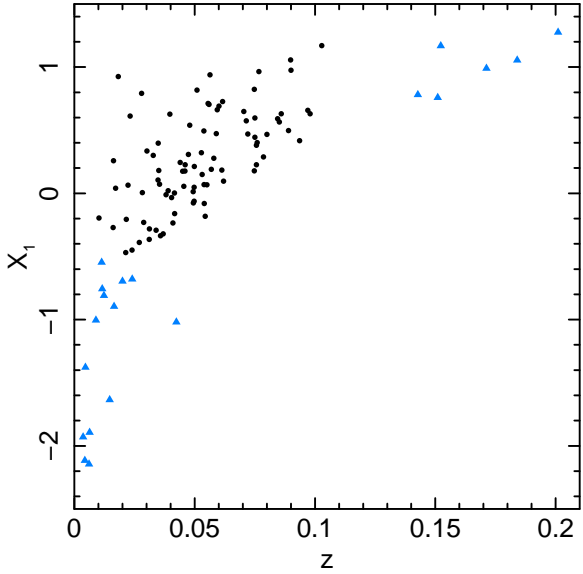}
  \caption{Luminosity--redshift distribution of the \RB{} data. The quantity $X_1$ is defined in \eqnref{eq:MLmodel2} and is computed in our reference cosmology. The data from \RB{} which are excluded from the analysis (see text), are shown as blue triangles.  The remaining clusters have an extent of roughly one decade in luminosity at all redshifts.}
  \label{fig:ML_z_X1}
\end{figure}

The process of fitting the model in \eqnref{eq:MLmodel} is complicated by the presence of Malmquist bias. Close to the flux limit for selection, any X-ray selected sample will preferentially include the most luminous sources for a given mass. This results in a steepening of the derived mass--luminosity relation and a bias in the inferred intrinsic scatter in luminosity for a given mass. (The inferred intrinsic dispersion may be artificially reduced or increased, depending on the distribution of data with respect to the flux limit.) The use of the extended sample of \RB{}, rather than only their flux-limited HIFLUGCS sample, partially mitigates this effect by softening the flux limit. Furthermore, we eliminate the 6 clusters at redshifts $>0.11$ which all lie close to the flux limit; as \figref{fig:ML_z_X1} shows, the extent in luminosity of the remaining 86 clusters is roughly a decade at all redshifts. As our estimate (see below) of the intrinsic scatter in $X_1$ is roughly 0.12 (\ie{} 0.12 decades in luminosity), and the Poisson measurement error is much smaller, we expect the effects of Malmquist bias on our estimation of the power-law slope and intrinsic dispersion to be minimal.

A second consequence of Malmquist bias is a strong apparent, but not necessarily physical, correlation between luminosity and redshift due to the fact that the flux limit corresponds to higher luminosities at higher redshifts. (This is evident in \figref{fig:ML_z_X1}.) Within redshift 0.11, we expect this false signal to be much greater than any real evolution; we therefore fix $\gamma=0$ when fitting the model.

Although methods exist for performing linear regression on data with bivariate heteroscedastic errors and intrinsic scatter \citep[e.g.][]{Akritas96}, such methods do not provide a simple goodness-of-fit measure that can be associated with arbitrary values of the fit parameters. In contrast, the \chisq{} statistic is an easily calculated measure of goodness-of-fit, but is biased by the presence of intrinsic scatter. We compromise by using a modified \chisq{} statistic which accounts for intrinsic scatter by introducing an additional dispersion term,
\begin{equation}
  \label{eq:modchisq}
  \modchisq = \sum_j \frac{(\alpha+\beta X_{1,j} - Y_j)^2}{\mscat_{Y,j}^2 + \delta},
\end{equation}
where $\mscat_{Y,j}$ is the measurement error on $Y_j$ and $\delta$ is the additional dispersion. This statistic is defined in terms of $Y$ given $X_1$ because the luminosity measurement errors are negligible in comparison to the mass measurement errors (\figref{fig:ML_x1_y}). The value of $\delta$ is chosen iteratively, adjusting it such that the best fit, found by minimizing \modchisq{} using least-squares methods, has \modchisq{} equal to the median of the chi-square distribution with $\nu=86-2$ (86 data points minus 2 free parameters) degrees of freedom (\ie{} $\modchisq/\nu\approx 1$). The \modchisq{} statistic thus does not measure absolute goodness-of-fit, but goodness relative to the best fit. Note that in the cosmological analysis we use the fixed value $\delta=0.014$ from our reference cosmology (see \secref{sec:MLlikelihood} for details).

A common log-normal intrinsic scatter in luminosity for a given mass can be estimated from the scatter in the data about the best fit, with measurement errors subtracted in quadrature. Since these data are predominantly at very low redshift, we use this procedure to estimate the $z=0$ value of the intrinsic dispersion (see \eqnref{eq:MLsevol}),
\begin{equation}
  \label{eq:MLs_estimation}
  \hat{\iscat}^2 = \frac{1}{\nu} \sum_{j} \left[ \left( \frac{Y_j-\alpha_f}{\beta_f} - X_{1,j} \right)^2 - \mscat_{X_1,j}^2 - \left(\frac{\mscat_{Y,j}}{\beta_f}\right)^2 \right],
\end{equation}
where $\alpha_f$ and $\beta_f$ are the parameters describing the best fit. Under assumptions of normal measurement errors and intrinsic scatter, the quantity $\hat{\iscat}^2\nu/\iscat^2$ is drawn from a chi-square distribution with $\nu$ degrees of freedom. This likelihood for the measured value $\hat{\iscat}$ can then be used to constrain the model parameter \iscat{}.

\begin{figure}
  \center
  \includegraphics[angle=270]{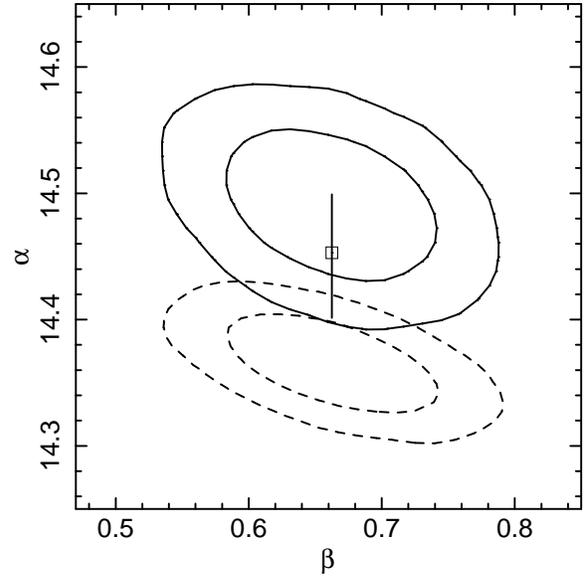}
  \caption{68.3 and 95.4 per cent confidence regions for the mass--luminosity parameters $\alpha$ and $\beta$ from the \RB{} data (solid lines). Also shown is the best fit and 68.3 per cent confidence interval (square with error bars) on $\alpha$ with fixed $\beta$ obtained from the \citet{Dahle06} data (\secref{sec:lensing}). The cosmology is fixed at our reference. The dashed lines are the confidence regions that would have been obtained from the \RB{} data if the masses were not corrected for bias due to the assumption of hydrostatic equilibrium. The difference in the precision of the $\alpha$ constraints between the corrected and uncorrected \RB{} results comes from our marginalization over the uncertainty in the size of the correction.}
  \label{fig:Dahle_RB_conf}
\end{figure}

The scatter in mass bias between clusters, \biasscatter{}, is propagated into the mass error bars, $\mscat_{Y}$, and thus impacts directly the estimated intrinsic scatter, $\hat{\iscat}$ (\eqnref{eq:MLs_estimation}). This results in a degeneracy between \biasscatter{} and \iscat{}; however, \biasscatter{} is not coupled to the other fit parameters, and so we can take its systematic uncertainty into account simply by increasing the width of the likelihood function for \iscat{}. We therefore replace the chi-square likelihood discussed above by the gamma distribution (of which the chi-square distribution is a special case). Fixing the location of the distribution's peak and augmenting its variance by a factor of $F$, the shape and scale parameters of the gamma distribution, $k$ and $\theta$, can be determined by solving the system of equations
\begin{eqnarray}
  \label{eq:chisq_gamma}
  (k-1)\theta & = & \nu - 2, \nonumber \\
  k \theta^2 & = & (2 \nu) F.
\end{eqnarray}
In practice, the value $F=14$ approximately corresponds to the desired 20 per cent uncertainty in \biasscatter{}. Finally, we account for a 20 per cent systematic uncertainty in the mean bias, \meanbias{}, by straightforwardly marginalizing over an appropriate prior. In detail, we parametrize the mean bias through $B=-\log_{10}(1-\meanbias)=\log_{10}(\mysub{M}{true}/\mysub{M}{est})$, since this quantity is added to the log-masses, $Y$.

\begin{figure}
  \center
  \includegraphics[angle=270]{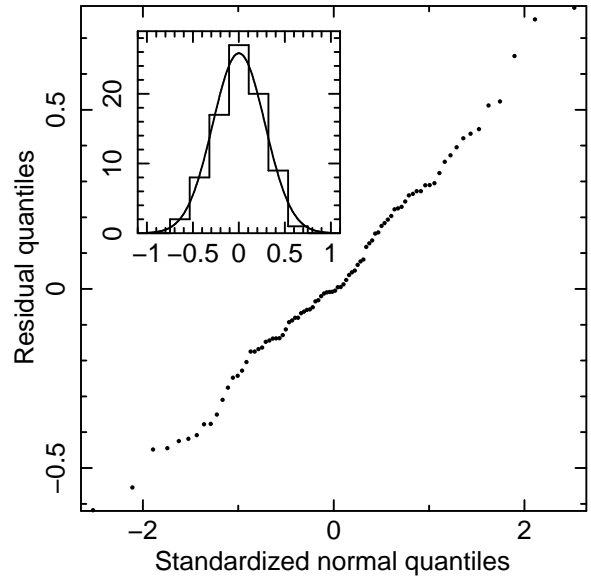}
  \caption{Quantile-quantile plot comparing the residuals in $X_1$ of the \RB{} data for the best fitting relation to the normal distribution. The straightness of the distribution indicates that the scatter is well approximated as normal in $X_1$ (log-normal in luminosity). (See footnote $^\thefootnote$.) Inset: the histogram of $X_1$ residuals is directly compared with the normal distribution.}
  \label{fig:RB_qqplot}
\end{figure}

Confidence regions for $\alpha$ and $\beta$ obtained with cosmological parameters fixed at our reference and using uniform priors on $\alpha$, $\beta$ and \iscat{} are displayed in \figref{fig:Dahle_RB_conf} (solid lines). The best-fitting values and marginal 68.3 per cent confidence intervals are $\alpha=14.49\pm 0.04$, $\beta=0.67^{+0.04}_{-0.06}$ and $\iscat=0.12^{+0.04}_{-0.03}$. A quantile-quantile plot of the residuals in $X_1$ for the best fit (\figref{fig:RB_qqplot}) confirms that the distribution of luminosities for a given mass is reasonably approximated by the log-normal distribution.\footnote{The quantile function for a probability distribution is the inverse of the cumulative distribution function (CDF). \figref{fig:RB_qqplot} shows each residual $r_i$ against the quantity $F^{-1}_N(F_E(r_i))$, where $F_N$ is the standard normal CDF and $F_E$ is the empirical CDF of the residuals. If $F_E$ is approximately normal, the points will be distributed roughly along a line. Significant curvature would indicate a poor match in distribution.}

\subsubsection{Weak lensing data}
\label{sec:lensing}

\begin{figure}
  \center
  \includegraphics[angle=270]{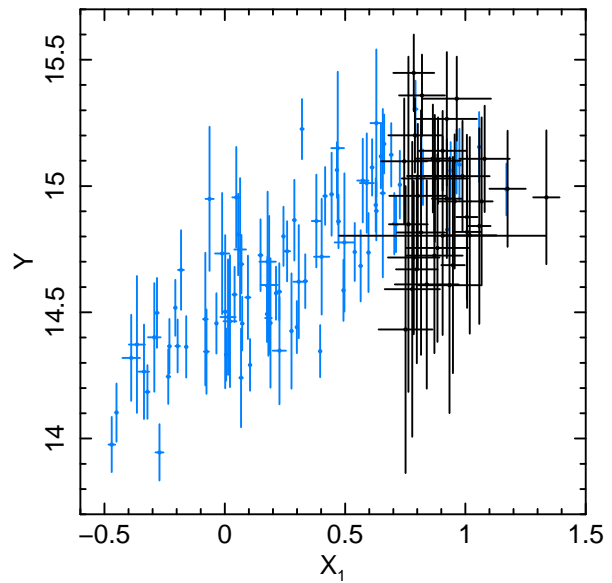}
  \caption{The mass--luminosity data of \citet{Dahle06} (black) are compared with those of \RB{} (blue; see additional comments of \figref{fig:ML_x1_y}). The effect of the (e)BCS flux limit on the Dahle data ($X_1\sim 0.75$) is clearly evident. At lower masses ($Y$), the data set is progressively more likely to contain only the most luminous clusters for that mass, resulting in an apparent steepening of the relation.}
  \label{fig:Dahle_x1_y}
\end{figure}

In order to verify the appropriateness of the hydrostatic bias correction applied to the \RB{} masses, we compare the results of \secref{sec:RBdata} with the data of \citet{Dahle06}, for which masses were measured using weak gravitational lensing. These masses, with associated luminosities taken from the BCS and extended BCS (eBCS) catalogs, are shown in \figref{fig:Dahle_x1_y} (black points). In contrast to the \RB{} data, the Malmquist bias due to the (e)BCS flux limit is clearly evident; in particular, the least massive clusters in this data set are very likely to represent the upper tail of the distribution of luminosities for given mass. Although the simple methods described above are unsuitable for fitting the mass--luminosity relation when the data are subject to significant Malmquist bias, the intersection of these two data sets near the highest observed masses indicates qualitatively that they are compatible.

We can quantify this observation by using the \citet{Dahle06} data to fit for the normalization, $\alpha$, by minimizing \modchisq{}, while leaving the slope, $\beta$, fixed at the best-fitting value from \secref{sec:RBdata}. The resulting one-dimensional 68.3 per cent confidence interval is $\alpha=14.46^{+0.04}_{-0.05}$ (\figref{fig:Dahle_RB_conf}), in agreement with the corrected \RB{} data. In contrast, the lensing constraint is incompatible at the 68 per cent confidence level with the confidence regions that would have been obtained from the X-ray data without applying the bias correction (dashed lines in \figref{fig:Dahle_RB_conf}).

\subsection{X-ray luminosity function}
\label{sec:XLF}

We use three flux-limited surveys in our analysis: the BCS \citep{Ebeling98} and REFLEX sample \citep{Bohringer04} at low redshifts ($z<0.3$), and the MACS \citep{Ebeling01} at $0.3<z<0.7$. (We restricted the REFLEX sample to the southern hemisphere so that its coverage on the sky would not overlap the BCS.) As discussed in Section~\ref{sec:XLFlikelihood}, a proper accounting of the intrinsic scatter in the mass--luminosity relation involves convolving over all possible masses when evaluating the likelihood of a set of cosmological parameters; it is therefore necessary for the mass--luminosity relation to be well calibrated at luminosities significantly (at least an order of magnitude) below those allowed in the surveys. We thus also restrict the analysis to clusters with large inferred luminosities $\hat{L}>2.55\E{44} h_{70}^{-2}\erg\second^{-1}$.

The completeness of the REFLEX sample was investigated by \citet{Bohringer01} and \citet{Schuecker01}, and is thought to be $>90$ per cent at a flux limit of $3.0\times 10^{-12} \erg\second^{-1} \cm^{-2}$. The BCS completeness as a function of flux is quantified in \citet{Ebeling98}; we use a flux limit of $4.4\times 10^{-12} \erg \second^{-1} \cm^{-2}$ where the BCS completeness is 90 per cent. Most of the incompleteness in the BCS is due to the inefficient extended-source detection of the RASS II algorithm, which is most severe at very low redshifts. One consequence of our high luminosity cut is that the sample contains mostly higher redshift objects ($\bar{z}\approx 0.21$) and not the large number of low redshift, low luminosity objects that would be included in a strictly flux-limited sample. The reported completeness of the BCS is thus a significant underestimate for the sub-sample of very luminous clusters used in our analysis. We have repeated the analysis of the BCS data alone (see \secref{sec:LCDMresults}) using a higher flux limit ($5\times 10^{-12} \erg \second^{-1} \cm^{-2}$) where the reported BCS incompleteness is negligible, and obtain a similar result. We conclude that significant incompleteness is not present in the BCS, given our selection criteria. The similar numbers of clusters (78 and 80, respectively) in the BCS and REFLEX samples satisfying the BCS flux limit and our luminosity cut also support this conclusion.

Unlike the BCS and REFLEX, for which extended-source fluxes were measured using the Voronoi tessellation and percolation \citep{Ebeling93,Ebeling93a} and growth curve analysis \citep{Bohringer00} algorithms, respectively, reported MACS fluxes are measured within a fixed angular size aperture (5 arcmin in most cases). A redshift-dependent correction for missing flux, described in \citet{Ebeling01}, is required when converting this aperture flux to total flux, both when determining luminosities for the detected clusters themselves and when computing the aperture flux-dependent survey sky coverage \citep{Ebeling07}. We adopt an aperture flux limit of $2\times 10^{-12} \erg \second^{-1} \cm^{-2}$, for which the completeness is $>90$ per cent \citep{Ebeling01}.

\begin{figure}
  \center
  \includegraphics[angle=270]{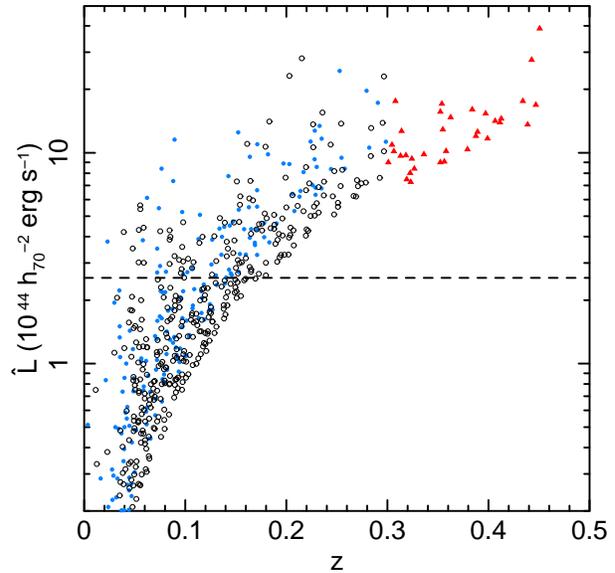}
  \caption{Luminosity--redshift distribution of clusters in the BCS (blue, filled circles), REFLEX (black, open circles) and MACS (red triangles) which are above the respective flux limits (see text). The adopted minimum luminosity of $2.55\E{44} h_{70}^{-2}\erg\second^{-1}$ is indicated by the dashed line. Error bars are not shown. Note that even though no clusters above our adopted flux limit were found at redshifts $0.5<z<0.7$, the lack of any such detections is significant in the analysis.}
  \label{fig:zLfig}
\end{figure}

Using the flux and redshift limits and the luminosity cut discussed above, the BCS, REFLEX and MACS samples respectively contribute 78, 130 and 34 clusters to our sample. The luminosity--redshift distribution of these data is displayed in \figref{fig:zLfig}.

In order to adequately take the effects of Eddington bias into account when predicting the number of clusters above our flux and luminosity thresholds for a given cosmology and mass--luminosity relation, we must be able to assign a probability to the luminosity observed from a cluster, given its true luminosity. Ultimately, the distribution is related to a Poisson distribution in the number of photons detected; however, the nontrivial conversion from photon count rate to unabsorbed flux and the variation in exposure times over the sky make the solution from first principles computationally difficult. Instead, we simplify the problem by assuming that the distribution $P(\hat{L}|L)$, where $\hat{L}$ and $L$ are the observed and true luminosities, is normal with mean $L$. The variance is a function of the observed flux, which we estimate empirically for each sample by fitting the flux errors versus flux to a power-law model. We find power-law slopes of 0.52 for the BCS and REFLEX samples and 0.56 for MACS, consistent with approximately Poisson scaling.

\subsection{Other data}
\label{sec:otherdata}

The primary purpose of this paper is to present an analysis based only on the X-ray luminosity function (XLF) data described above, along with the priors described in \secref{sec:analysis}. However, we also show results obtained from other cosmological data for comparison purposes, along with results of a combined analysis, in \secref{sec:results}. Specifically, the additional data consist of observations of galaxy cluster gas mass fractions (\fgas{}), the cosmic microwave background (CMB), and type Ia supernovae (SNIa). The SNIa and \fgas{} results shown are identical to those in \citet{Allen07}. The \fgas{} data are reported in that work and the SNIa results are derived from the compilation of \citet{Davis07}, which includes data from the ESSENCE survey \citep[60 targets;][]{Wood-Vasey07,Miknaitis07}, the SNLS first year data \citep[57 targets;][]{Astier06}, 45 nearby supernovae \citep{Jha07} and the 30 high-redshift supernovae discovered by the Hubble Space Telescope and reported by \citet{Riess07} for which a `gold' rating was awarded (192 SNIa in total). Our analysis of the cosmic microwave background anisotropies uses three-year Wilkinson Microwave Anisotropy Probe (WMAP) data, including marginalization over a plausible range in the amplitude of the Sunyaev-Zel'dovich signal in galaxy clusters ($0<\mysub{A}{SZ}<2$) \citep{Spergel07}. We use the October 2006 version of the WMAP likelihood code.\footnote{http://lambda.gsfc.nasa.gov}

When analysing the CMB data, as with the XLF analysis, we include the effects of density perturbations in the dark-energy fluid. For our standard analysis, we fix the dark-energy sound speed in units of the speed of light, $c_s$, to 1.0. (Some discussion of the influence of perturbations can be found in \secref{sec:soundspeed}.) In addition, no prior on the scalar spectral index, $n_s$, is used in the CMB analysis, since WMAP places very strong constraints on this parameter ($0.951\pm0.016$; \citealt{Spergel07}); as noted in \secref{sec:priors}, this degree of variation in $n_s$ produces negligible change in the XLF results.

\section{Analysis}
\label{sec:analysis}

We parametrize the full model fitted to the X-ray luminosity function data as $\{h$, $\Omegab h^2$, $\Omegac h^2$, $\sigma_8$, $n_s$, $w$, $A$, $\alpha$, $\beta$, $\gamma$, $\iscat$, $\iscatprime$, $B\}$, where \Omegab{} and \Omegac{} are the baryon and cold dark matter densities ($\Omegam=\Omegab+\Omegac$), $A$ is the normalization of the Jenkins mass function (\eqnref{eq:jmf}), and the last 6 parameters describe the mass--luminosity relation and its calibration. In addition to the assumption of spatial flatness, we adopt the Gaussian priors $h=0.72\pm 0.08$ \citep{Freedman01} and $\Omegab h^2=0.0214\pm 0.002$ \citep{Kirkman03} from the Hubble Key Project and Big Bang nucleosynthesis studies, respectively. These latter priors are necessary because the likelihood depends very weakly on $h$ and $\Omegab$, which enter only through the transfer function. We additionally marginalize over a 20 per cent uncertainty in the normalization, $A$, of the \JMF{} mass function to account for the residuals of the fitting formula to their simulations over the mass range of interest and the expected variation among cosmologies. We must also place a prior on the mass--luminosity evolution parameters $\gamma$ and \iscatprime{}, which are not constrained by our analysis of the \RB{} data. For the standard set of allowances used in this paper, we adopt the uniform priors $|\gamma|<0.35$, which corresponds to a limit of approximately 20 per cent evolution in the normalization of the mass--luminosity relation within redshift 0.7, and $|\iscatprime|<0.3$. Since the results are insensitive to the spectral index within a reasonable range (\secref{sec:priors}), we fix $n_s=0.95$ in accordance with \citet{Spergel07} for the standard analysis. The dark-energy equation of state was bounded by a uniform prior, $-5<w<0$. For the remaining model parameters, uniform priors on the physically allowed domains were used. These priors are summarized in \tabref{tab:priors} (labeled ``standard'' priors). The sensitivity of our results to the choice of priors is analyzed in \secref{sec:priors}.

\begin{table}
  \caption{Priors used in the analysis. All parameters not listed were assigned uniform priors on their physically allowed domains. $N(\mu,\sigma)$ indicates the normal distribution with mean $\mu$ and variance $\sigma^2$, and $U(a,b)$ indicates the uniform distribution with endpoints $a$ and $b$.
  \label{tab:priors}}
  \centering
  \begin{tabular}{lcl}
    Prior & Parameter & Density \\
    \hline
    standard: & $w$ & $U(-5.0,0.0)$ \\
    & $h$ & $N(0.72,0.08)$ \\
    & $\Omegab h^2$ & $N(0.0214,0.002)$ \\
    & $n_s$ & fixed at 0.95 \\
    & $c_s^2$ & fixed at 1.0 \\
    & $A$ & $N(0.316,0.0632)$ \\
    & $\gamma$ &  $U(-0.35,0.35)$ \\
    & \iscatprime{} & $U(-0.3,0.3)$ \\
    & $B$ & $N(-0.125,0.03)$ \\
    & \biasscatter{} &  See \secref{sec:RBdata} \vspace{1mm} \\
    weak: & $h$ &  $N(0.72,0.16)$ \\
    & $\Omegab h^2$ &  $N(0.0214,0.004)$ \\
    & $n_s$ & $N(0.95,0.05)$ \\
    & $c_s^2$ & $U(0.0,1.0)$ \\
    & $A$ & $N(0.316,0.1264)$ \\
    & $\gamma$ &  $U(-0.7,0.7)$ \\
    & \iscatprime{} & $U(-0.6,0.6)$ \\
    & $B$ & $N(-0.125,0.06)$ \\
    \hline
  \end{tabular}
\end{table}

Cosmological constraints were determined via Markov Chain Monte Carlo, employing the Metropolis algorithm; the calculations were performed using a modified version of the COSMOMC code\footnote{http://cosmologist.info/cosmomc/} of \citet[][see also \citealt{Rapetti05,Rapetti07}]{Lewis02}. Multiple, randomly initialized Markov chains were produced for each set of results, and convergence to the posterior distribution was monitored using the Gelman-Rubin criterion, $R$, which measures the ratio of between-chain to within-chain variances \citep{Gelman92}, as well as by visual inspection. Acceptable convergence was defined by the requirement $R-1<0.05$. The log-likelihood of the data given a set of model parameters is decomposed into the sum $\ell_{\alpha\beta} + \ell_\iscatsym + \llxlf$, whose terms are described in the remainder of this Section.

\subsection{Mass--luminosity likelihood}
\label{sec:MLlikelihood}

The parameters describing the normalization and slope of the mass--luminosity relation, $\alpha$ and $\beta$, were constrained using the \modchisq{} statistic defined in \eqnref{eq:modchisq}. Only the \RB{} data were used. As mentioned previously, we fix the value of $\delta$ to be 0.014, calculated in our reference cosmology using the method described in \secref{sec:RBdata}. This makes values of \modchisq{} from different steps of the Markov chain directly comparable, while retaining the de-biasing effect of the modification to \chisq. (The reference cosmology with respect to which \modchisq{} is defined is unimportant, since the Markov chain is sensitive only to differences in the statistic.) The log-likelihood associated with \modchisq{} is defined as $\ell_{\alpha\beta} = -\modchisq/2$.

At each step of the chain, the estimated intrinsic dispersion in the mass--luminosity relation, $\hat{\iscat}^2$,  is computed by \eqnref{eq:MLs_estimation}. The contribution to the log-likelihood from the dispersion is the logarithm of the gamma distribution described in \secref{sec:RBdata},
\begin{equation}
  \label{eq:ll_sigma}
  \ell_\iscatsym = \left(k-1\right)\ln\left(x\right) - \frac{x}{\theta} - k\ln\left(\theta\right),
\end{equation}
where $k$ and $\theta$ are the solutions to \eqnref{eq:chisq_gamma} for $\nu=84$ and $F=14$, $x=\nu\hat{\iscat}^2/\iscat^2$, and where we have omitted a normalization term that does not depend on the model parameters. This likelihood penalizes models for which the intrinsic dispersion is far from the estimated dispersion measured from the \RB{} data.

\subsection{Luminosity function likelihood}
\label{sec:XLFlikelihood}

The likelihood that $N$ clusters with inferred luminosities in a range $d\hat{L}$ exist in a volume $dV$ can in general be written as a Poisson probability plus a correction due to the clustering of halos with one another. Given that our sample covers a very wide survey area ($\sim 2/3$ of the sky) and includes only the most luminous, and therefore rare, objects, the clustering term is negligible compared to the pure Poisson term \citep[e.g.][]{Hu03}. If the plane of redshift and inferred luminosity is divided into non-overlapping cells, then the likelihood of our data is simply
\begin{equation}  \label{eq:xlflike}
  P\left(N_1,N_2,\ldots\right) = \prod_j \frac{\tilde{N}_j^{N_j}e^{-\tilde{N}_j}}{N_j!},
\end{equation}
where $N_j$ and $\tilde{N}_j$ are the number of clusters detected and predicted in the $j$th cell, respectively. The log-likelihood is then
\begin{equation}
  \label{eq:xlfloglike}
  \llxlf = \sum_j \left[ N_j\ln(\tilde{N}_j) - \tilde{N}_j \right],
\end{equation}
where the model-independent term $-\sum_j \ln(N_j!)$ has been dropped.

If the cells are taken to be rectangular, with the $j$th cell given by $z_j^{(1)} \leq z < z_j^{(2)}$ and $\hat{L}_{j}^{(1)} \leq \hat{L} < \hat{L}_{j}^{(2)}$, then
\begin{equation}
  \label{eq:modelN}
  \tilde{N}_j = \int_{z_j^{(1)}}^{z_j^{(2)}}dz \frac{dV(z)}{dz} \int_{\hat{L}_{j}^{(1)}}^{\hat{L}_{j}^{(2)}}d\hat{L} \frac{d\tilde{n}(z,\hat{L})}{d\hat{L}},
\end{equation}
where $V(z)$ is the comoving volume within redshift $z$. In the absence of intrinsic scatter in the mass--luminosity relation and measurement errors in the observed luminosities, the derivative of the comoving number density would be simply
\begin{equation}
  \label{eq:dndLsimple}
  \frac{d\tilde{n}(z,L)}{dL} = \fsky(z,L) \frac{dM_{500}(L)}{dL} \frac{dn(z,M_{324m})}{dM_{324m}} \frac{dM_{324m}}{dM_{500}}.
\end{equation}
Here \fsky{} is the sky coverage fraction of the surveys as a function of redshift and inferred luminosity (\ie{} flux), $dn/dM_{324m}$ is the Jenkins mass function (\eqnref{eq:mfcn_def}) and $M_{500}(L)$ is the mass--luminosity relation (\eqnref{eq:MLpowlaw}). The presence of scatter requires us to take into account that a cluster detected with inferred luminosity $\hat{L}$ could potentially have any true luminosity $L$ and mass $M_{500}$, with some associated probability. To calculate the predicted number density correctly, we must therefore convolve with these probability distributions:
\begin{eqnarray}
  \label{eq:dndL}
  \frac{d\tilde{n}(z,\hat{L})}{d\hat{L}} & = & \fsky(z,\hat{L}) \int_0^\infty dL ~ P(\hat{L}|L) \\ 
  & \times & \int_0^\infty dM_{500} ~ P(L|M_{500}) \frac{dn(z,M_{324m})}{dM_{324m}} \frac{dM_{324m}}{dM_{500}}.\nonumber
\end{eqnarray}
Above, $P(L|M_{500})$ is a log-normal distribution whose width is the intrinsic scatter in the mass--luminosity relation, $\iscatsym{}(z)$, and $P(\hat{L}|L)$ is a normal distribution whose width as a function of flux is modeled as a power law, as described in \secref{sec:XLF}. When evaluating the conversion between $M_{500}$ and $M_{324m}$ using the method of \citet{Hu03}, we assume a universal concentration of 4.0 at radius $r_{200}$, consistent with the results of \citet{Gao07}, who find that the concentrations of very massive clusters varies little with mass and redshift through $z=1$. We have verified that including a log-normal scatter in concentration of $\sim 0.15$ \citep[e.g.][]{Comerford07,Schmidt07} has no effect on our results.

The sum over the second term in \eqnref{eq:xlfloglike} reduces to the integrated number of predicted clusters in the detection region of the surveys (\ie{} within the redshift range and above the luminosity and flux thresholds), independent of binning. However, the first term in that equation is dependent on the choice of binning. To make optimal use of the data the bin size should be very small, in which case the integrals in \eqnref{eq:modelN} can be approximated as the integrand multiplied by $\Delta z \Delta \hat{L}$. By introducing a factor of $d_L^2/d_L^2=1$, where $d_L$ is the luminosity distance to the mean redshift of the bin, we can factor out the constant and cosmologically-invariant bin area $\Delta z (\Delta \hat{L} / d_L^2)$; the logarithm of this term in \eqnref{eq:xlfloglike} is independent of model parameters and may thus be dropped, simplifying the log-likelihood to
\begin{equation}
  \llxlf = \sum_i \ln\left( d_L^2 \left. \frac{dV}{dz}\frac{d\tilde{n}}{d\hat{L}} \right|_{z_i,\hat{L}_i} \right) + \int dz d\hat{L} \frac{dV}{dz} \frac{d\tilde{n}}{d\hat{L}},
\end{equation}
where the summation is over detected clusters and the integral extends over the detection region of the surveys.

\section{Results}
\label{sec:results}

\newcommand*{\stdprior}{1}
\newcommand*{\hWbweakprior}{2}
\newcommand*{\nsweakprior}{3}
\newcommand*{\Aweakprior}{4}
\newcommand*{\gweakprior}{5}
\newcommand*{\pweakprior}{6}
\newcommand*{\Bweakprior}{7}
\newcommand*{\Bsfixedprior}{8}
\newcommand*{\Bzeroprior}{9}
\newcommand*{\Fscatprior}{10}
\newcommand*{\MLscatprior}{11}
\newcommand*{\csprior}{12}
\newcommand*{\nopertprior}{13}
\begin{table*}
  \begin{minipage}{175mm}
    \caption{Best-fitting values and 68.3 per cent confidence intervals for the model parameters obtained from the luminosity function data. Our main results from this study occupy the first two lines; the remaining results are listed in the order that they appear in the text. $^a$B=BCS, R=REFLEX, M=MACS. $^b$Priors not specified below are standard (see \tabref{tab:priors}). \stdprior: all standard; \hWbweakprior: weak $h$ and $\Omega_b h^2$; \nsweakprior: weak $n_s$; \Aweakprior: weak $A$; \gweakprior: weak $\gamma$; \pweakprior: weak \iscatprime{}; \Bweakprior: weak B; \Bsfixedprior: \meanbias{} and \biasscatter{} fixed at nominal values; \Bzeroprior: \meanbias{} and \biasscatter{} fixed to 0; \Fscatprior: fixed $\iscat=0$; \MLscatprior: survey luminosity measurement error$=0$; \csprior: marginalized over $c_s^2$; \nopertprior: no dark energy density perturbations.
      \label{tab:constraints}}
    \centering
    \begin{tabular}{ccccccccc}
      Data$^a$ & Model & Priors$^b$ & \Omegam{} & $\sigma_8$ & $w$ & $\alpha$ & $\beta$ & $\iscat$ \vspace{3mm}  \\
      \hline
      B+R+M & \LCDM & \stdprior & $0.28^{+0.11}_{-0.07}$ & $0.78^{+0.11}_{-0.13}$ & --- & $14.49\pm 0.04$ & $0.67^{+0.03}_{-0.05}$ & $0.15^{+0.14}_{-0.05}$ \\
      B+R+M & \wCDM & \stdprior & $0.24^{+0.15}_{-0.07}$ & $0.85^{+0.13}_{-0.20}$ & $-1.4^{+0.4}_{-0.7}$ & $14.48\pm 0.04$ & $0.65^{+0.05}_{-0.04}$ & $0.15^{+0.13}_{-0.05}$ \vspace{3mm} \\
      B     & \LCDM & \stdprior & $0.26^{+0.25}_{-0.09}$ & $0.78^{+0.10}_{-0.37}$ & --- & $14.51\pm 0.04$ & $0.67\pm 0.05$ & $0.18^{+0.18}_{-0.09}$ \vspace{0.5mm} \\
      R     & \LCDM & \stdprior & $0.20^{+0.10}_{-0.04}$ & $0.85^{+0.10}_{-0.09}$ & --- & $14.49\pm 0.04$ & $0.66^{+0.04}_{-0.05}$ & $0.12^{+0.05}_{-0.03}$ \vspace{0.5mm} \\
      M     & \LCDM & \stdprior & $0.30^{+0.24}_{-0.10}$ & $0.73^{+0.14}_{-0.13}$ & --- & $14.50\pm 0.04$ & $0.67^{+0.04}_{-0.05}$ & $0.12^{+0.05}_{-0.03}$ \vspace{0.5mm} \\
      B+R+M & \wCDM & \hWbweakprior & $0.25^{+0.19}_{-0.9}$ & $0.78^{+0.18}_{-0.19}$ & $-1.4^{+0.5}_{-0.7}$ & $14.48\pm 0.04$ & $0.65\pm 0.05$ & $0.16^{+0.13}_{-0.06}$ \vspace{0.5mm} \\
      B+R+M & \wCDM & \nsweakprior & $0.24^{+0.17}_{-0.08}$ & $0.86^{+0.10}_{-0.24}$ & $-1.5^{+0.5}_{-0.6}$ & $14.48\pm 0.04$ & $0.65\pm 0.05$ & $0.16^{+0.14}_{-0.06}$ \vspace{0.5mm} \\
      B+R+M & \wCDM & \Aweakprior & $0.25^{+0.21}_{-0.08}$ & $0.83^{+0.13}_{-0.21}$ & $-1.4^{+0.5}_{-1.0}$ & $14.48\pm 0.04$ & $0.65\pm 0.05$ & $0.16^{+0.14}_{-0.06}$ \vspace{0.5mm} \\
      B+R+M & \wCDM & \gweakprior & $0.24^{+0.15}_{-0.07}$ & $0.85^{+0.12}_{-0.20}$ & $-1.5^{+0.6}_{-0.8}$ & $14.48\pm 0.04$ & $0.65\pm 0.05$ & $0.15^{+0.13}_{-0.05}$ \vspace{0.5mm} \\
      B+R+M & \wCDM & \pweakprior & $0.24^{+0.14}_{-0.07}$ & $0.84^{+0.13}_{-0.17}$ & $-1.5\pm 0.5$ & $14.48\pm 0.04$ & $0.66^{+0.04}_{-0.05}$ & $0.15^{+0.12}_{-0.05}$ \vspace{0.5mm} \\
      B+R+M & \wCDM & \Bweakprior & $0.24^{+0.17}_{-0.07}$ & $0.81^{+0.16}_{-0.17}$ & $-1.4^{+0.5}_{-0.7}$ & $14.48^{+0.08}_{-0.06}$ & $0.65\pm 0.05$ & $0.16^{+0.14}_{-0.06}$ \vspace{0.5mm} \\
      B+R+M & \wCDM & \Bsfixedprior & $0.22^{+0.06}_{-0.05}$ & $0.89\pm 0.09$ & $-1.4^{+0.3}_{-0.5}$ & $14.48\pm 0.03$ & $0.65^{+0.05}_{-0.04}$ & $0.118^{+0.010}_{-0.009}$ \vspace{0.5mm} \\
      B+R+M & \wCDM & \Bzeroprior & $0.21^{+0.07}_{-0.04}$ & $0.75^{+0.10}_{-0.05}$ & $-1.4^{+0.3}_{-0.5}$ & $14.34\pm 0.03$ & $0.66\pm 0.04$ & $0.205^{+0.013}_{-0.018}$ \vspace{0.5mm} \\
      B+R+M & \wCDM & \Fscatprior & $0.21^{+0.06}_{-0.05}$ & $0.93\pm 0.09$ & $-1.4^{+0.3}_{-0.4}$ & $14.48\pm 0.04$ & $0.65^{+0.04}_{-0.05}$ & --- \vspace{0.5mm} \\
      B+R+M & \wCDM & \MLscatprior & $0.21^{+0.08}_{-0.05}$ & $0.91^{+0.09}_{-0.10}$ & $-1.4^{+0.3}_{-0.4}$ & $14.48\pm 0.04$ & $0.67^{+0.04}_{-0.05}$ & $0.12^{+0.04}_{-0.03}$ \vspace{0.5mm} \\
      B+R+M & \wCDM & \csprior & $0.23^{+0.16}_{-0.06}$ & $0.85^{+0.12}_{-0.21}$ & $-1.5^{+0.5}_{-0.6}$ & $14.48\pm 0.04$ & $0.65\pm 0.05$ & $0.16^{+0.15}_{-0.06}$ \vspace{0.5mm} \\
      B+R+M & \wCDM & \nopertprior & $0.25^{+0.15}_{-0.08}$ & $0.85^{+0.10}_{-0.20}$ & $-1.6^{+0.7}_{-0.8}$ & $14.47\pm 0.04$ & $0.64\pm 0.05$ & $0.15^{+0.13}_{-0.05}$ \\
      \hline
    \end{tabular}
  \end{minipage}
\end{table*}

\subsection{\LCDM{} constraints}
\label{sec:LCDMresults}

\begin{figure}
  \center
  \includegraphics{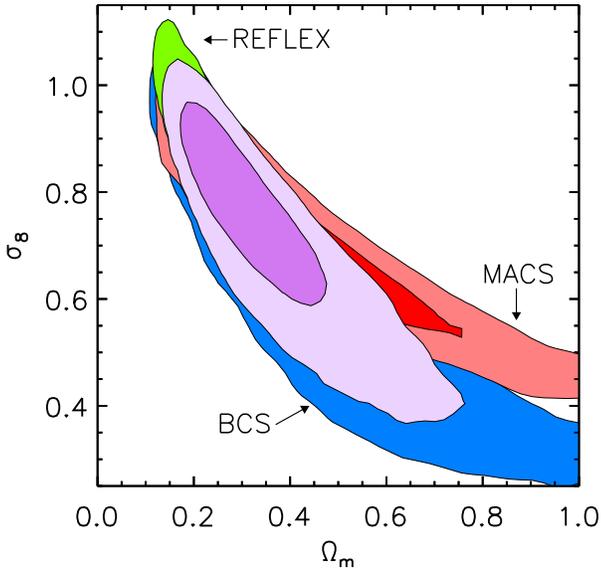}
  \caption{Joint 68.3 and 95.4 per cent confidence constraints on \Omegam{} and $\sigma_8$ for a \LCDM{} model from MACS (red), BCS (blue), and REFLEX (green) individually, and their combination (purple) using standard priors (\tabref{tab:priors}). Note that only the 95.4 per cent confidence regions are visible for the individual BCS and REFLEX data sets.}
  \label{fig:fLCDM_brm_compare}
\end{figure}

For a \LCDM{} cosmology, we compare the joint \Omegam-$\sigma_8$ constraints obtained from the BCS, REFLEX and MACS data sets individually, as well as their combination, in \figref{fig:fLCDM_brm_compare}. The marginalized constraints from the combination of the three cluster samples are $\Omegam=0.28^{+0.11}_{-0.07}$ and $\sigma_8=0.78^{+0.11}_{-0.13}$ (\tabref{tab:constraints}).

\begin{figure}
  \center
  \includegraphics{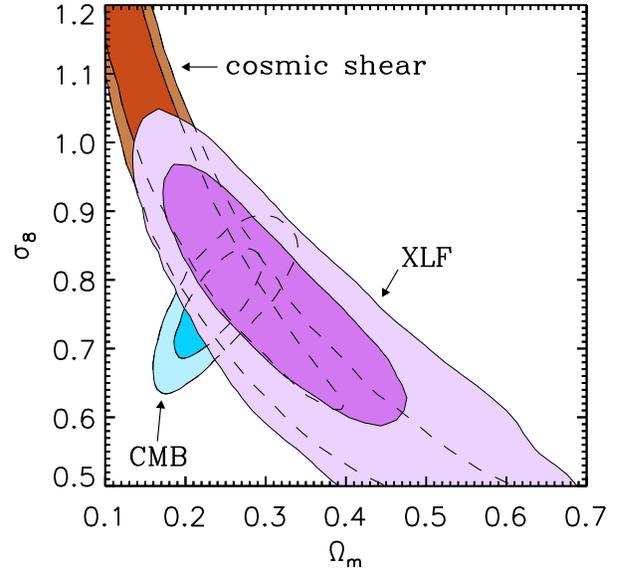}
  \caption{Joint 68.3 and 95.4 per cent confidence constraints on \Omegam{} and $\sigma_8$ for a \LCDM{} model using the combined X-ray luminosity function (XLF) data (purple) and our standard priors (\tabref{tab:priors}). Also shown are independent constraints from the CMB \citep[blue;][]{Spergel07} and cosmic shear \citep[brown;][]{Benjamin07}).}
  \label{fig:fLCDM_Wm_s8_compare}
\end{figure}

These constraints are in good agreement with recent, independent results from the CMB \citep{Spergel07} and cosmic shear, as measured in the 100 Square Degree Survey \citep{Benjamin07} (\figref{fig:fLCDM_Wm_s8_compare}) and CFHTLS Wide field \citep{Fu07}. Our results are also in good overall agreement with previous findings based on the observed X-ray luminosity and temperature functions of clusters \citep[e.g.][]{Eke98,Donahue99,Henry00,Borgani01,Seljak02,Allen03,Pierpaoli03,Schuecker03,Henry04}, although the correction to the hydrostatic mass estimates employed in the present study leads to our result on $\sigma_8$ being, typically, somewhat higher for a given value of \Omegam{} (see \secref{sec:massbias}). Our result on \Omegam{} is in excellent agreement with current constraints based on cluster \fgas{} data (\citealt{Allen07} and references therein) and the power spectrum of galaxies in the 2dF galaxy redshift survey \citep{Cole05} and Sloan Digital Sky Survey (SDSS) \citep{Eisenstein05,Tegmark06,Percival07}, as well as the combination of CMB data with a variety of external constraints \citep{Spergel07}. Our result on $\sigma_8$ is marginally lower than that determined by weak lensing tomography in the Cosmic Evolution Survey \citep[COSMOS;][]{Massey07} and by the observed number density of optically-selected groups and clusters in the 2dF \citep{Eke06} and SDSS surveys \citep{Rozo07}.

\subsection{\wCDM{} constraints}

\subsubsection{Results using the XLF data alone}
\label{sec:wCDMresults}

\begin{figure*}
  \center
  \includegraphics{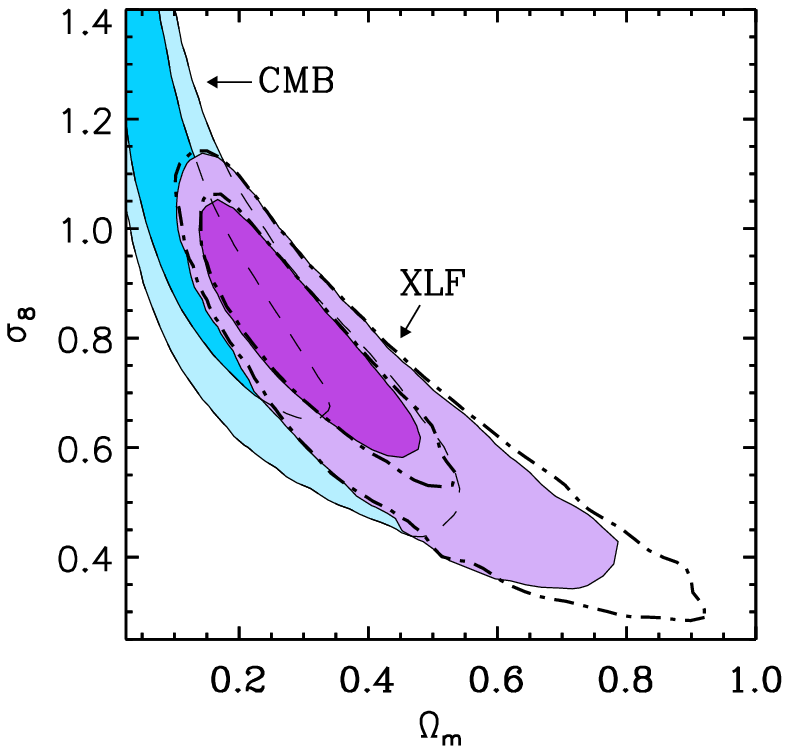}
  \hspace{11mm}
  \includegraphics{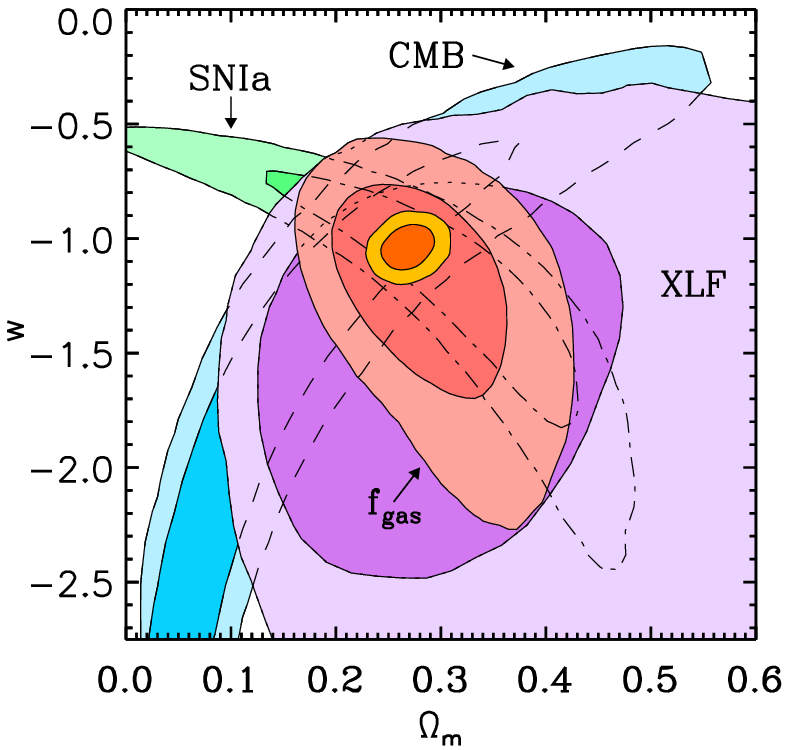}
  \caption{Joint 68.3 and 95.4 per cent confidence constraints on \Omegam{} and $\sigma_8$ (left) and \Omegam{} and $w$ (right) for a constant-$w$ model using the X-ray luminosity function data (purple) and standard priors (\tabref{tab:priors}). Also shown are independent constraints from the CMB \citep[blue;][]{Spergel07}, SNIa data \citep[green;][]{Davis07} and cluster \fgas{} data \citep[red][]{Allen07}, and the combination of all four (gold). In the left panel, the dot-dashed lines indicate the XLF results using our weak prior on $n_s$.}
  \label{fig:fwCDM_compare}
\end{figure*}

\begin{figure*}
  \center
  \includegraphics{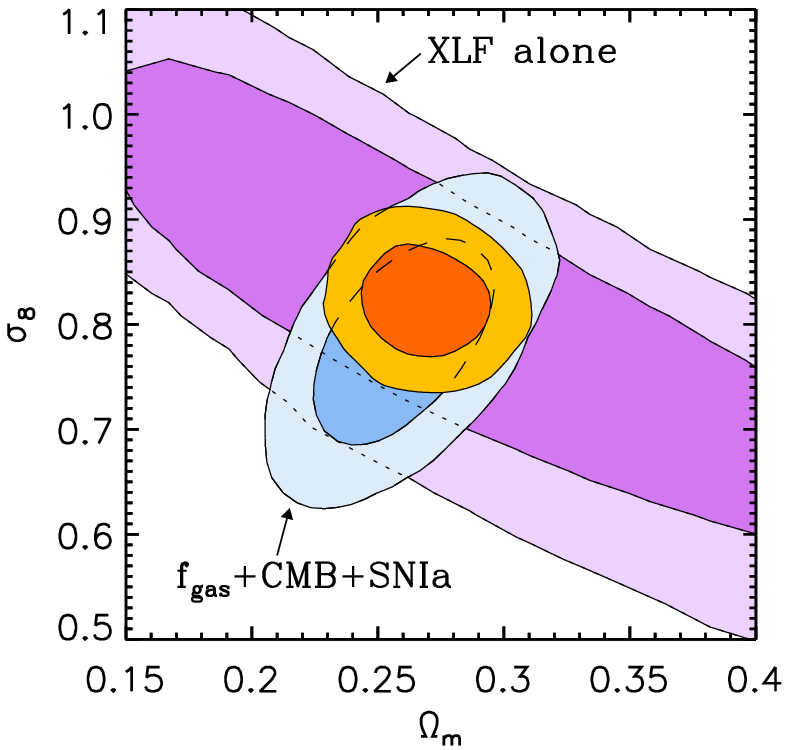}
  \hspace{11mm}
  \includegraphics{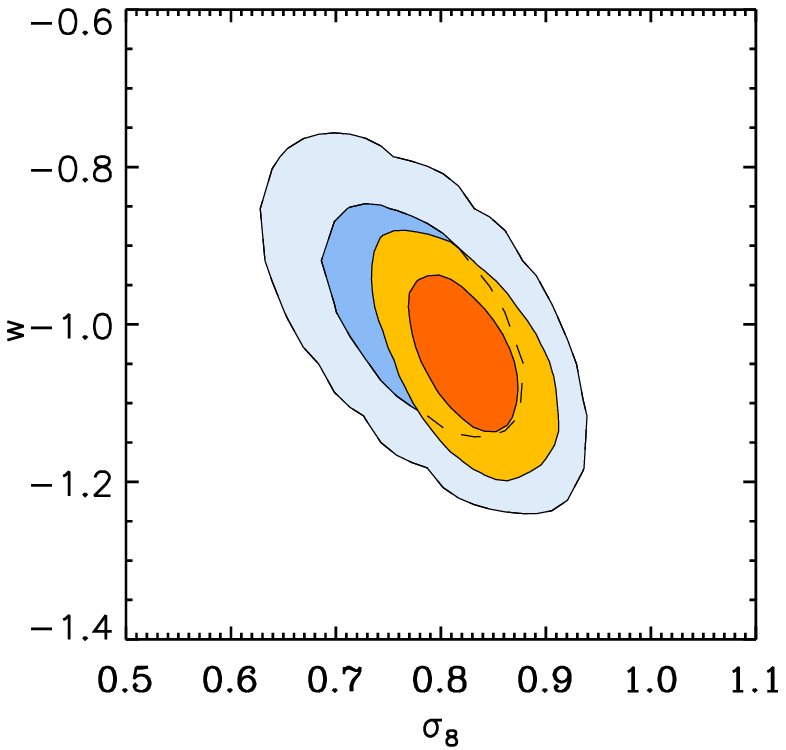}
  \caption{Joint 68.3 and 95.4 per cent confidence constraints on \Omegam{} and $\sigma_8$ (left) and $\sigma_8$ and $w$ (right) obtained from a combined \fgas{}+CMB+SNIa analysis (blue) and the improved constraints obtained by combining these data with the XLF (gold). No priors on $h$, $\Omegab h^2$ or $n_s$ are imposed in either analysis. In the left panel, the results from the XLF alone using standard priors (\tabref{tab:priors}) are shown (purple) in order to illustrate the degeneracy breaking.}
  \label{fig:combination}
\end{figure*}

The joint constraints on \Omegam{} and $\sigma_8$ from the luminosity function data using our standard priors (purple contours) are compared with those of WMAP (blue) in the left panel of \figref{fig:fwCDM_compare}. Since the WMAP results are marginalized over $n_s$, we also show the XLF results using our weak $n_s$ prior as dot-dashed lines, although the difference from the standard results is small. The right panel displays constraints on \Omegam{} and $w$ obtained independently from the XLF (purple), WMAP (blue), SNIa data (green) and cluster \fgas{} data (red). Our new XLF results are consistent with each of these independent data sets, and with the cosmological-constant model ($w=-1$). The marginalized results from the X-ray luminosity function data are $\Omegam=0.24^{+0.15}_{-0.07}$, $\sigma_8=0.85^{+0.13}_{-0.20}$ and $w=-1.4^{+0.4}_{-0.7}$ (\tabref{tab:constraints}).

\subsubsection{Combined XLF+\fgas{}+CMB+SNIa results}
\label{sec:combination}

We now consider the improvement over a combined \fgas{}+CMB+SNIa analysis \citep[following][]{Allen07} that can be achieved by additionally including the XLF data. The results of this comparison are displayed in \figref{fig:combination} and \tabref{tab:combination}. The \fgas{}+CMB+SNIa combination already provides tight constraints on \Omegam{}, $h$, $\Omegab h^2$ and $n_s$ (hence no priors on these parameters are used in either combined analysis), but the degeneracy between $w$ and $\sigma_8$ (right panel of \figref{fig:combination}) limits the precision of the dark energy results. The addition of the XLF data breaks the degeneracy in the \Omegam{}-$\sigma_8$ plane (left panel), resulting in tighter constraints on \Omegam{}, $\sigma_8$ and $w$.

\begin{table}
  \caption{Best-fitting values and marginalized 68.3 per cent confidence limits on cosmological parameters obtained from combined analyses of the \fgas{}+WMAP+SNIa and XLF+\fgas{}+WMAP+SNIa data. No external priors on $h$, $\Omegab h^2$ and $n_s$ are used.}
    \label{tab:combination}
  \centering
  \begin{tabular}{crr}
    & \fgas{}+WMAP & XLF+\fgas{} \\
    & +SNIa       & +WMAP+SNIa \\
    \hline
    \Omegam{} & $0.258\pm 0.022$ & $0.269\pm 0.016$ \\
    $\sigma_8$ & $0.79\pm 0.06$ & $0.82\pm 0.03$ \\
    $w$ & $-0.99\pm 0.07$ & $-1.02\pm 0.06$ \\
    \hline
  \end{tabular}
\end{table}

\subsection{Goodness of fit}

\begin{figure}
 \center
 \includegraphics{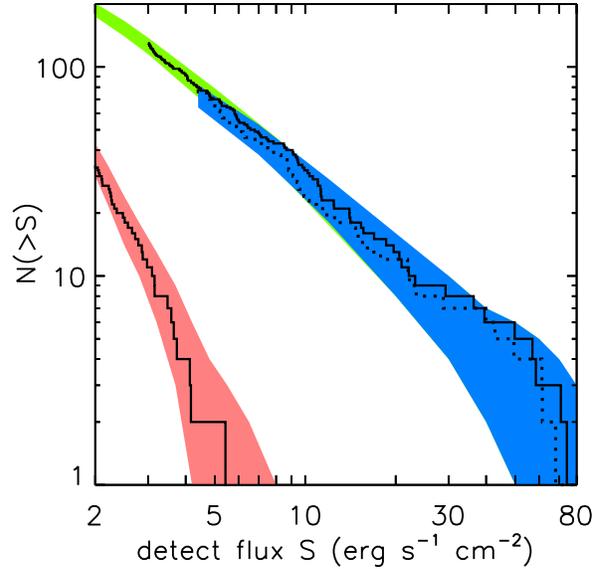}
 \caption{The number of objects $N$ above detect flux $S$ as a function of $S$ ($\log N-\log S$ distribution) for BCS (dotted line, above/right), REFLEX (solid line, above/right) and MACS (solid line, below/left). Shaded areas indicate the predictions and 68 per cent confidence regions of the best-fitting \LCDM{} model. Due to the similar sky areas covered, the BCS and REFLEX predictions are essentially identical at $S>8$.}
 \label{fig:goodness}
\end{figure}

In order to assess the goodness of fit for the XLF data, we compare the number of clusters in each sample with the number predicted by the best-fitting set of parameters in the Markov chains. Since the chains were generated at finite temperature, these do not necessarily correspond to the posterior modes, but they should be nearby. Using the XLF data only, the best-fitting \LCDM{} parameters predict 73, 127 and 37 clusters in BCS, REFLEX and MACS, respectively, while the best fitting \wCDM{} parameters predict 76, 130 and 35. Both are in good agreement with the true samples, which respectively contain 78, 130 and 34 clusters above flux and luminosity thresholds of \secref{sec:XLF}. This agreement for the \LCDM{} case is displayed graphically as a function of the flux limit in \figref{fig:goodness}. (The numbers quoted above correspond to the low-flux ends of the lines in the figure.) The equivalent figure for \wCDM{} looks similar.

\begin{figure}
  \center
  \includegraphics{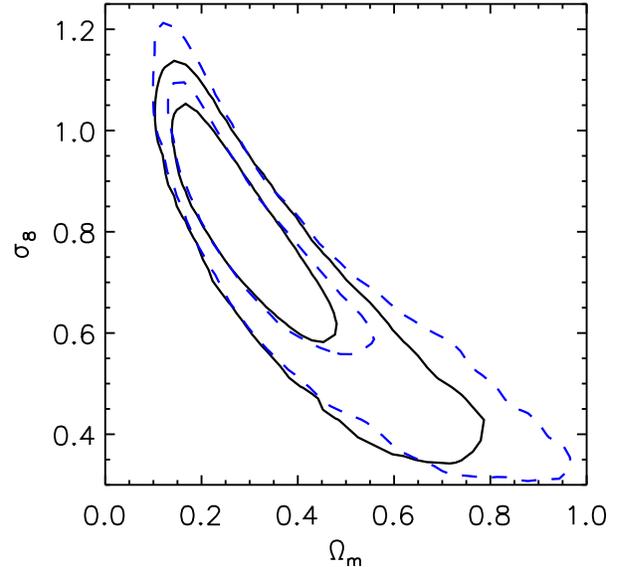}
  \caption{Comparison of joint 68.3 and 95.4 per cent confidence constraints on \Omegam{} and $\sigma_8$ for a constant-$w$ model using standard priors (solid, black lines) and weak priors on $h$ and $\Omegab h^2$ (dashed, blue lines). Weakening these priors results in a slight expansion of the confidence region along the \Omegam{}-$\sigma_8$ degeneracy axis. The priors have no affect on the determination of $w$.}
  \label{fig:priors_hWb_s8}
\end{figure}

\section{Discussion}
\label{sec:discussion}

\subsection{Sensitivity to priors}
\label{sec:priors}

The standard priors used in our analysis are conservative; however, we have also performed analyses using the even more conservative priors labeled as ``weak'' in \tabref{tab:priors}. These weak-prior results effectively demonstrate the degeneracies between parameters of interest (\Omegam{}, $\sigma_8$ and $w$) and each nuisance parameter, given the current level of statistical and systematic uncertainty. Two of these weak priors, on \iscatprime{} and $B$, did not result in significant changes to any of the cosmological constraints. The results of the other tests are summarized below. One-dimensional, marginalized results for each of these analyses are also presented in \tabref{tab:constraints}.

\subsubsection{Hubble constant and baryon density}
We consider $h$ and $\Omegab h^2$ together, since both influence the results only through the shape of the power spectrum. The results obtained by doubling the width of the Gaussian priors on these parameters (blue, dashed contours) is compared with our standard-prior constraint in \figref{fig:priors_hWb_s8}. The effect of weakening these priors is primarily to expand the constraints along the \Omegam{}-$\sigma_8$ degeneracy curve. The constraint on $w$ is not affected.

\subsubsection{Scalar spectral index}
We next consider the effect of marginalizing over $n_s$ using a Gaussian prior of width 0.05, significantly larger than the constraint obtained from WMAP \citep{Spergel07}. Like $h$ and $\Omegab h^2$, the effect of freeing this parameter is to weaken the constraints along the \Omegam{}-$\sigma_8$ degeneracy, although the effect is relatively small, even with this wide prior (left panel of \figref{fig:fwCDM_compare}).

\subsubsection{Mass function normalization}
\begin{figure}
  \center
  \includegraphics{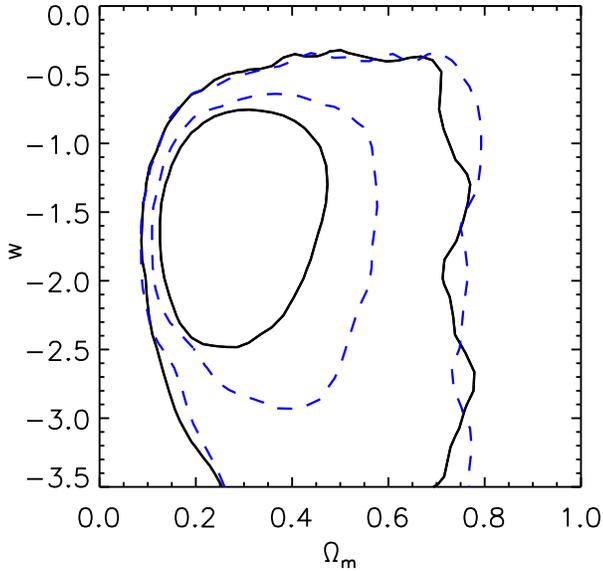}
  \caption{Comparison of joint 68.3 and 95.4 per cent confidence constraints on \Omegam{} and $w$ for a constant-$w$ model using standard priors (solid, black lines) and a weak prior on the normalization of the mass function, $A$ (dashed, blue lines). The constraints on \Omegam{} and $w$ are weakened, although $\sigma_8$ is not greatly affected (\tabref{tab:constraints}).}
  \label{fig:priors_A_w}
\end{figure}

The weak prior on $A$ doubles the uncertainty on the \JMF{} normalization to 40 per cent. The result is a decrease in constraining power on \Omegam{} and $w$, but not significantly on $\sigma_8$. The expansion of the dark-energy constraints can be seen in \figref{fig:priors_A_w}.

\subsubsection{Evolution in the mass--luminosity normalization}
Since $\gamma$ describes only the redshift evolution of the mass--luminosity relation, doubling the size of its uniform prior has no effect on the determination of the redshift-zero quantities \Omegam{} and $\sigma_8$, as can be seen in \tabref{tab:constraints}. However, there is a degeneracy with $w$, resulting in a poorer constraint (\tabref{tab:constraints}, prior \gweakprior{}).

\subsection{AGN contamination}
\label{sec:agn}

The presence of active galactic nuclei (AGN) and other point-like X-ray emitters is a potential concern for the accuracy of the mass--luminosity relation. There is no possibility of subtracting these point sources from the RASS data, since the number of photons in a cluster detection is typically too small. However, they are subtracted from the \RB{} luminosities, to the extent that \ROSAT{} can resolve them. In addition, there is the possibility of an increase in the density of AGN over the redshift range of our data (e.g. \citealt*{Hasinger05} and references therein). There may therefore be an inequivalence between luminosities inferred from the RASS data and those used in the calibration of the mass--luminosity relation. In our work studying MACS clusters (in preparation), we take advantage of the \Chandra{} X-ray Observatory's high spatial resolution to efficiently identify point sources and determine their contribution to the total cluster luminosity. We find that this contribution is at the few per cent level -- much smaller than the typical survey flux uncertainty -- and conclude that point source contamination is not an issue for the current study.

\begin{figure*}
  \center
  \includegraphics{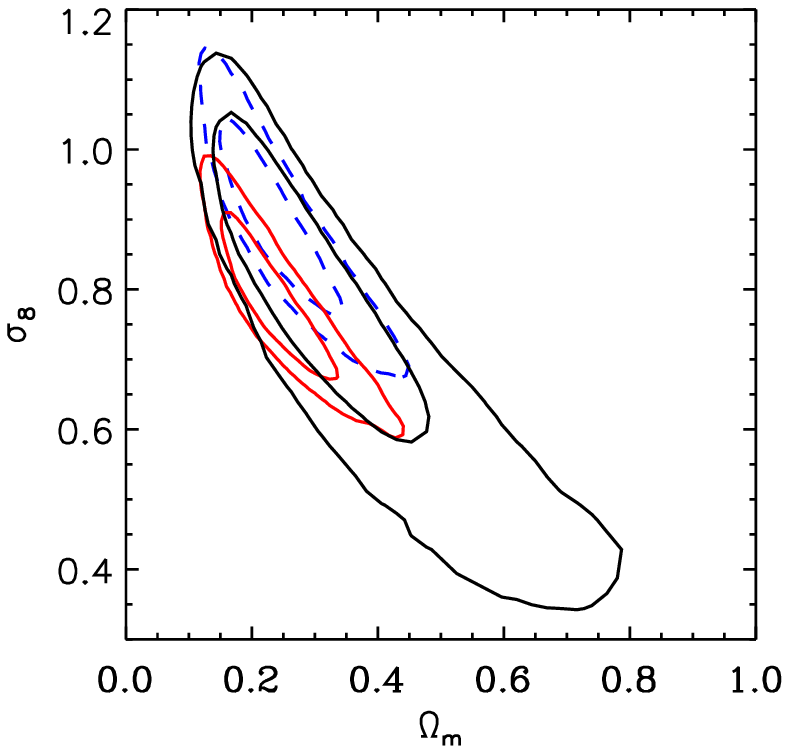}
  \hspace{11mm}
  \includegraphics{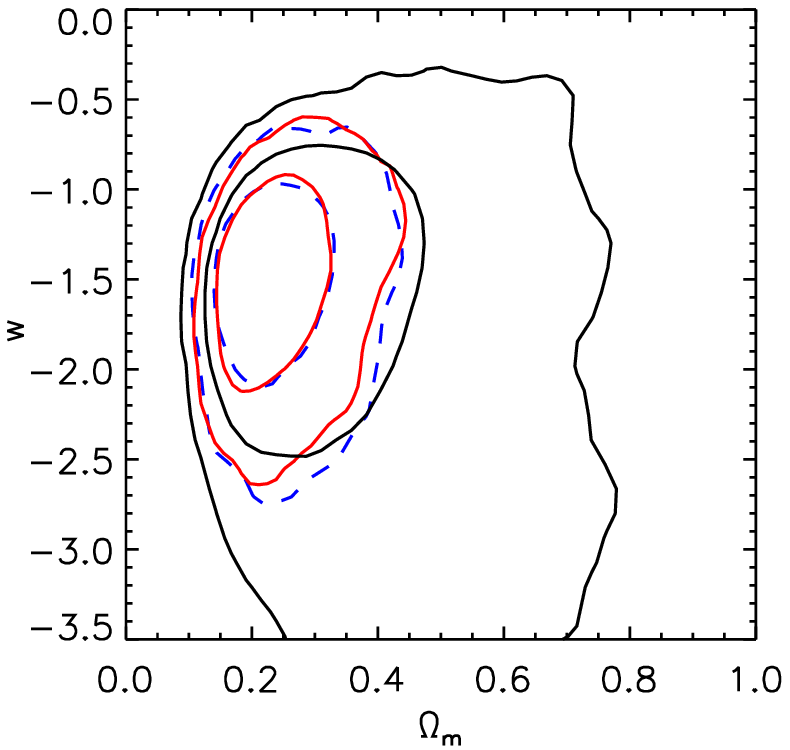}
  \caption{Comparison of joint 68.3 and 95.4 per cent confidence constraints for a constant-$w$ model using standard priors (black, solid lines), results with \meanbias{} and \biasscatter{} fixed at their nominal values (blue, dashed lines), and results with no correction for hydrostatic mass bias ($\meanbias{}=0,\biasscatter{}=0$; red, solid lines). The uncertainty in \biasscatter{} in particular has a significant impact on our results (see text). Neglecting the uncertainties on \meanbias{} and \biasscatter{} produces spuriously tight constraints on \Omegam{}, $\sigma_8$ and $w$. Neglecting completely the hydrostatic mass bias results in a significantly lower estimate of $\sigma_8$.}
  \label{fig:massbias}
\end{figure*}

\begin{figure*}
  \center
  \includegraphics{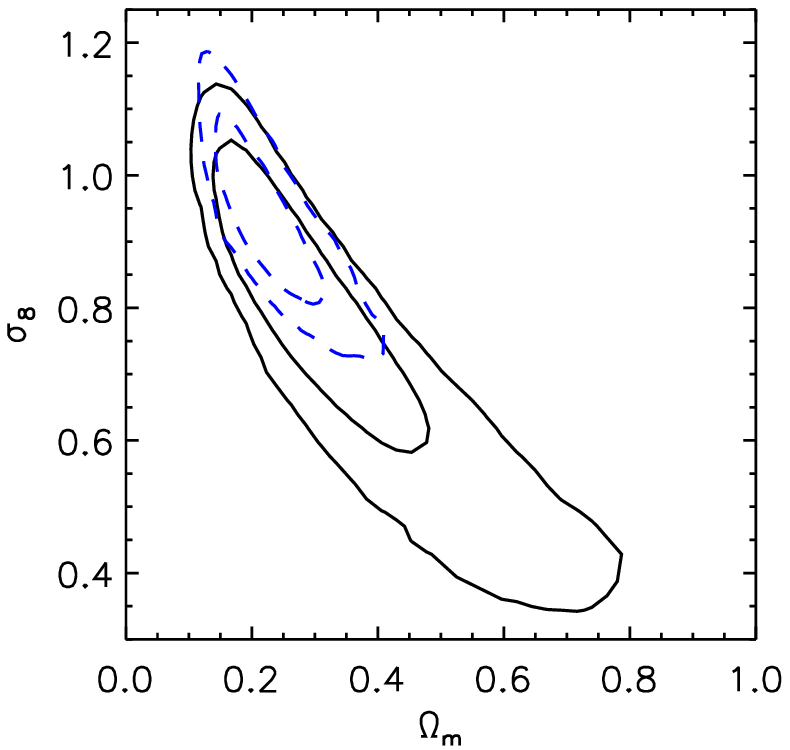}
  \hspace{11mm}
  \includegraphics{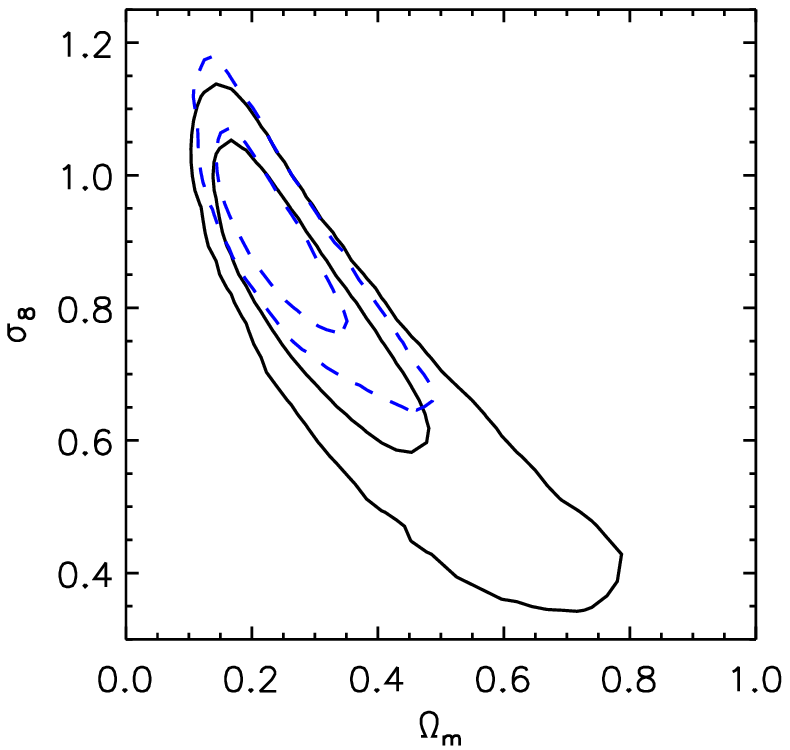}
  \caption{Comparison of joint 68.3 and 95.4 per cent confidence constraints on \Omegam{} and $\sigma_8$ for a constant-$w$ model using standard priors (solid, black lines) and results obtained by ignoring either the intrinsic dispersion in the mass--luminosity relation (left) or the survey flux measurement errors (right) (dashed, blue lines). Failing to account for the intrinsic dispersion in the mass--luminosity relation produces a significant bias towards higher values of $\sigma_8$ and lower values of \Omegam{}; a similar, but smaller, bias is evident when the survey flux measurement error is not accounted for.}
  \label{fig:MLFscat}
\end{figure*}

\subsection{Bias in mass measurements}
\label{sec:massbias}

The bias in the measured masses of the mass--luminosity data is completely degenerate with the mass function normalization, $A$, which in turn affects inferences made on the other parameters. Consequently, it is important to include in the analysis the mean correction to masses motivated by departures from hydrostatic equilibrium, \meanbias{}, and the scatter in this bias among clusters, \biasscatter{}, as well as the uncertainty in both quantities. \figref{fig:massbias} shows that the constraints obtained by fixing the mean bias and bias scatter at their nominal values (blue, dashed contours) are spuriously tighter than those of our standard analysis (black, solid contours) by $\sim 50$ per cent on \Omegam{} and $\sigma_8$ and $\sim 30$ per cent on $w$. Also shown is the result of neglecting the bias correction entirely (red, solid contours), which results in lower values of $\sigma_8$ compared with the nominal correction.

Comparing the results for priors \Bweakprior{} and \Bsfixedprior{} in \tabref{tab:constraints} with our standard-prior results, it is clear that the uncertainty in \biasscatter{} has a greater effect than the uncertainty in \meanbias{}. This is encouraging, since the former is due primarily to the very small number of simulations used to calibrate the observational mass bias. Straightforwardly increasing the number of simulated clusters should result in significant improvement to our constraints.

\subsection{Intrinsic and measurement luminosity scatter}
\label{sec:MLFscat}

The mass distribution of clusters available to a flux-limited survey is influenced by the degree of scatter in flux for a given mass. As described in \secref{sec:XLFlikelihood}, this effect can be decomposed into convolutions due to intrinsic scatter in the mass--luminosity relation and measurement error in the survey flux determinations. Failure to account for either of these sources of scatter when evaluating the number of detectable clusters predicted by a set of model parameters can significantly bias the result. The magnitude of the effect is demonstrated in  \figref{fig:MLFscat}, which shows the results for a \wCDM{} cosmology obtained by ignoring either of these scatters individually. In each case, the results are biased towards lower \Omegam{} and higher $\sigma_8$, along the degeneracy between the two parameters. The constraints on cosmological parameters, including $w$, are also spuriously tight in both cases. (The joint \Omegam-$w$ constraints are not shown, but closely resemble the tighter results in \figref{fig:massbias}.) This results from the fact that any scatter in luminosity compounds the systematic effect of uncertainty in \meanbias{}. In the case where intrinsic mass--luminosity dispersion is ignored, the contributions due to uncertainty in \iscat{}, \iscatprime{} and \biasscatter{} also vanish from the error budget.

\subsection{Dark-energy perturbations}
\label{sec:soundspeed}

In the analysis thus far, we have accounted for dark-energy density perturbations assuming that the sound speed is equal to the speed of light ($c_s^2=1$). \tabref{tab:constraints} (prior \csprior) shows that marginalizing over such a constant sound speed ($0<c_s^2<1$) has no effect on our results (but note that dark-energy models generally predict a time-varying sound speed).

\citet{Weller03}, \citet{Rapetti05} and \citet{Spergel07} showed that including such perturbations greatly reduces the ability of WMAP data to discriminate among models with $w<-1$ (\figref{fig:fwCDM_compare}, \cf{} Figures 15 and 16 of Spergel et al.). In contrast, the effect of dark-energy perturbations on the XLF constraints is less dramatic; \tabref{tab:constraints} (prior \nopertprior{}) shows that an analysis with no perturbations results in somewhat larger constraints on $w$, opposite in character from the WMAP analysis.

\section{Conclusion}
\label{sec:conclusion}

We have presented new constraints on cosmological-constant (\LCDM{}) and constant-$w$ (\wCDM{}) dark-energy models using the observed X-ray luminosity function of the largest, most X-ray luminous galaxy clusters out to redshift 0.7, in combination with standard priors on $h$ and $\Omegab h^2$. At 68.3 per cent confidence, we find $\Omegam=0.28^{+0.11}_{-0.07}$ and $\sigma_8=0.78^{+0.11}_{-0.13}$ for a \LCDM{} model, and $\Omegam=0.24^{+0.15}_{-0.07}$, $\sigma_8=0.85^{+0.13}_{-0.20}$ and $w=-1.4^{+0.4}_{-0.7}$ for a \wCDM{} model. These results include marginalization over uncertainties in the theoretical mass function, non-self-similar evolution in the mass--luminosity relation, and the bias in mass estimates from X-ray observations. Our results constitute the first precise determination of the dark-energy equation of state using measurements of the growth of cosmic structure observed in galaxy clusters, and provide additional support for the cosmological-constant model.

These results build upon, and are largely in agreement with, a number of earlier galaxy cluster studies (see \secref{sec:introduction}). Our constraints on cosmological parameters are consistent with independent findings from studies of type Ia supernovae, anisotropies in the CMB, the X-ray gas mass fraction of galaxy clusters, galaxy redshift surveys and leading cosmic shear surveys. The agreement between the results from these independent techniques is reassuring, and motivates a combined analysis of the data in order to investigate more complex models of dark energy. Combining our data with CMB, SNIa and \fgas{} data, we find $\Omegam=0.269\pm 0.016$, $\sigma_8=0.82\pm 0.03$ and $w=-1.02\pm 0.06$.

The results for $\sigma_8$ presented here are somewhat higher than those from previous work based on the BCS and REFLEX data due to our correction to the masses used to constrain the mass--luminosity relation. The magnitude of this discrepancy underscores the need for an improved understanding of the observational biases resulting from asphericity, projection effects and hydrostatic disequilibrium. More advanced and comprehensive simulations, calibrated by gravitational lensing studies, show considerable promise in this area. Broad-band, high spectral resolution X-ray data from, for example, the New X-ray Telescope (NeXT) and Constellation-X, will allow precise measurements of gas velocities and non-thermal emission components in clusters, providing a more comprehensive understanding of the relevant gas physics. In the short term, simply increasing the number of simulated clusters used to calibrate the observational bias would reduce the uncertainty in the scatter in bias among clusters. Since the uncertainty in this scatter is the dominant source of systematic uncertainty in $\sigma_8$, this approach provides a relatively easy means of improving the precision of our results.

With regard to dark energy, the systematics due to the theoretical mass function and the redshift evolution of the mass--observable relation are also important. The former can be addressed with a large suite of cosmological simulations, allowing a more precise parametrization of the cosmological dependence of the mass function. The latter necessitates a more rigorous study of galaxy cluster virial relations and their evolution, carefully accounting for selection effects. Such improved analysis techniques will be required if future, high-redshift X-ray (e.g. Spectrum-RG/eROSITA) or Sunyaev-Zel'dovich surveys are to be used to their full potential.

\section*{Acknowledgments}
\label{sec:acknowledgements}

We thank Adrian Jenkins for providing computer code to evaluate the mass function of dark matter halos and Gil Holder, Alexey Vikhlinin and Jeremy Tinker for helpful discussions. We also thank Glenn Morris and Stuart Marshall for computer support. Calculations for this work were carried out using the KIPAC XOC and Orange compute clusters at the Stanford Linear Accelerator Center (SLAC) and the SLAC Unix compute farm. We acknowledge support from the National Aeronautics and Space Administration (NASA) through NASA LTSA grant NAG5-8253, and though Chandra Award Numbers DD5-6031X, GO2-3168X, GO2-3157X, GO3-4164X, GO3-4157X and GO5-6133, issued by the Chandra X-ray Observatory Center, which is operated by the Smithsonian Astrophysical Observatory for and on behalf of NASA under contract NAS8-03060. This work was supported in part by the U.S. Department of Energy under contract number DE-AC02-76SF00515. AM was additionally supported in part by a William~R. and Sara Hart Kimball Stanford Graduate Fellowship.

While this work was in revision, the 5-year WMAP results were released (\citealt{Dunkley08} and references therein). The new WMAP constraints are also in excellent agreement with our results. Since the difference between the 3-year and 5-year WMAP results is only marginal, we have not repeated our combined analysis with the newer CMB data in this work.

\bibliographystyle{mnras}
\def \aap {A\&A}
\def \statisci {Statis. Sci.}
\def \pre {Phys.\ Rev.\ E}
\def \apj {ApJ}
\def \apjl {ApJL}
\def \apjs {ApJS}
\def \mnras {MNRAS}
\def \prd {Phys.\ Rev.\ D}
\def \prl {Phys.\ Rev.\ Lett.}

\bsp
\label{lastpage}
\end{document}